\documentclass[fleqn,usenatbib]{mnras}

\usepackage{newtxtext} 
\usepackage{mathptmx}
\usepackage{txfonts}

\usepackage[T1]{fontenc}
\usepackage{xcolor} 
\DeclareRobustCommand{\VAN}[3]{#2}
\let\VANthebibliography\thebibliography
\def\thebibliography{\DeclareRobustCommand{\VAN}[3]{##3}\VANthebibliography}

\usepackage{graphicx}	

\usepackage{xfrac}

\DeclareRobustCommand{\ion}[2]{%
\relax\ifmmode
\ifx\testbx\f@series
{\mathbf{#1\,\mathsc{#2}}}\else
{\mathrm{#1\,\mathsc{#2}}}\fi
\else\textup{#1\,{\mdseries\textsc{#2}}}%
\fi}

\title[REBELS: The Infrared Luminosity Function at $\mathrm{z \sim 7}$]{The ALMA REBELS Survey: The First Infrared Luminosity Function Measurement at $\mathbf{z \sim 7}$}

\author[L. Barrufet et al.]{L. Barrufet$^{1}$\thanks{E-mail:laia.barrufetdesoto@unige.ch}, 
P. A. Oesch$^{1,2}$,
R. Bouwens$^{3}$,
H. Inami$^{4}$,
L. Sommovigo$^{5}$,
H. Algera$^{4,6}$,
E. da Cunha$^{7}$, 
\and
M. Aravena$^{8}$,
P. Dayal$^{9}$,
A. Ferrara$^{5}$,
Y. Fudamoto$^{10,11}$,
V. Gonzalez$^{12,13}$
L. Graziani$^{14,15}$
A. Hygate$^{3}$,
\and
I. de Looze$^{16,17}$,
T. Nanayakkara$^{18}$,
A. Pallottini$^{5}$,
R. Schneider$^{14, 15}$
M. Stefanon$^{19, 20}$
M. Topping$^{21}$  
\and
and P. van Der Werf$^{3}$
\\
$^{1}$Department of Astronomy, University of Geneva, Chemin Pegasi 51, 1290 Versoix, Switzerland \\
$^{2}$Cosmic Dawn Center (DAWN), Niels Bohr Institute, University of Copenhagen, Jagtvej 128, K\o benhavn N, DK-2200, Denmark\\
$^{3}$Leiden Observatory, Leiden University, PO Box 9500, 2300 RA Leiden, The Netherlands  \\
$^{4}$ Hiroshima Astrophysical Science Center, Hiroshima University, 1-3-1, Kagamiyama-Higashi, Hiroshima 739-8526, Japan \\
$^{5}$ Scuola Normale Superiore, Piazza dei Cavalieri, 7, 56126 Pisa, Italy \\
$^{6}$ National Astronomical Observatory of Japan, 2-21-1, Osawa, Mitaka, Tokyo, Japan  \\
$^{7}$ International Centre for Radio Astronomy Research, University of Western Australia, Stirling Hwy, Crawley, 26WA, 6009, Australia \\
$^{8}$ Nucleo de Astronomia Facultad de Ingenieria y Ciencias, Universidad Diego Portales, Av Ejercito 441, Santiago, Chile \\
$^{9}$ Kapteyn Astronomical Institute, University of Groningen, P.O. Box 800, 9700, AV Groningen, The Netherlands \\
$^{10}$ Waseda Research Institute for Science and Engineering, Faculty of Science and Engineering, Waseda University, 3-4-1 Okubo, Shinjuku, Tokyo 169-8555, Japan \\
$^{11}$National Astronomical Observatory of Japan, 2-21-1, Osawa, Mitaka, Tokyo, Japan \\
$^{12}$ Departmento de Astronomia, Universidad de Chile, Casilla 36-D, Santiago 7591245, Chile \\
$^{13}$Centro de Astrofisica y Tecnologias Afines (CATA), Camino del Observatorio 1515, Las Condes, Santiago 7591245, Chile \\
$^{14}$ Dipartimento di Fisica, Sapienza, Universita di Roma, Piazzale Aldo Moro 5, I-00185 Roma, Italy \\
$^{15}$ INAF/Osservatorio Astronomico di Roma, via Frascati 33, I-00078 Monte Porzio Catone, Roma, Italy
\\
$^{16}$ Sterrenkundig Observatorium, Ghent University, Krijgslaan 281-59, 9000, Gent, Belgium \\
$^{17}$ Dept. of Physics \& Astronomy, University College London, Gower Street, London WC1E 6BT, United Kingdom \\
$^{18}$ Centre for Astrophysics \& Supercomputing, Swinburne University of Technology, PO Box 218, Hawthorn, VIC 3112, Australia \\
$^{19}$ Departament d'Astronomia i Astrofisica, Universitat de Valencia, C. Dr. Moliner 50, E-46100 Burjassot, Valencia,  Spain \\
cia, C. Dr. Moliner 50, E-46100 Burjassot, Valencia,  Spain \\
$^{20}$ Unidad Asociada CSIC "Grupo de Astrofisica Extragalactica y Cosmologia" (Instituto de Fisica de Cantabria - Universitat de Valencia) \\
$^{21}$  Steward Observatory University of Arizona, 933 N Cherry Ave, Tucson, AZ 85721, United States \\
}
\date{Accepted XXX. Received YYY; in original form ZZZ}

\pubyear{2022}

\begin{document}
\label{firstpage}
\pagerange{\pageref{firstpage}--\pageref{lastpage}}
\maketitle

\begin{abstract}
We present the first observational infrared luminosity function (IRLF) measurement in the Epoch of Reionization (EoR) based on a UV-selected galaxy sample with ALMA spectroscopic observations. Our analysis is based on the ALMA large program Reionization Era Bright Emission Line Survey (REBELS), which targets 42 galaxies at $\mathrm{z=6.4-7.7}$ with [CII] 158$\micron$ line scans. 16 sources exhibit a dust detection, 15 of which are also spectroscopically confirmed through the [CII] line. The IR luminosities of the sample range from $\log L_{IR}/L_\odot=11.4$ to 12.2. Using the UVLF as a proxy to derive the effective volume for each of our target sources, we derive IRLF estimates, both for detections and for the full sample including IR luminosity upper limits. The resulting IRLFs are well reproduced by a Schechter function with the characteristic luminosity of $\log L_{*}/L_\odot=11.6^{+0.2}_{-0.1}$. Our observational results are in broad agreement with the average of predicted IRLFs from simulations at $z\sim7$. Conversely, our IRLFs lie significantly below lower redshift estimates, suggesting a rapid evolution from $z\sim4$ to $z\sim7$, into the reionization epoch. The inferred obscured contribution to the cosmic star-formation rate density at $z\sim7$ amounts to $\mathrm{log(SFRD/M_{\odot}/yr/Mpc^{3}) = -2.66^{+0.17}_{-0.14} }$ which is at least $\sim$10\% of UV-based estimates. We conclude that the presence of dust is already abundant in the EoR and discuss the possibility of unveiling larger samples of dusty galaxies with future ALMA and JWST observations. 
\end{abstract}


\newcommand{\TBD}[1]{\textcolor{red}{\bf{#1}}}	

\section{Introduction}
\label{Introduction}

It is still a crucial open question in astrophysics when the first galaxies formed and how they built up their mass. The continuous discovery of higher redshift galaxies is pushing the boundaries of our knowledge of galaxy evolution \citep[e.g.,][]{Dunlop2013_review, Stark16, Dayal2018, Schaerer2022,  Naidu2022_Schrodinger, Atek2022,  Adams2022}. In particular,  the discovery of a significant population of luminous and massive galaxies at $\mathrm{z > 9}$ has posed questions about the speed of early stellar mass production \citep[e.g.][]{Oesch2016, Laporte2021, Naidu2022_z11_z13, Labbe2022}. 

Until recently, the knowledge of galaxies at $\mathrm{z>7}$ was mainly based on rest-frame ultraviolet (UV) observations \citep{Oesch2018, Bouwens2021}. These samples might not be complete, however, as they might miss extremely dust obscured, but highly star-forming galaxies \citep[e.g.,][]{Casey2019}.

From an observational point of view, the Atacama Large Millimeter Array (ALMA) is the most powerful tool to study dust at high redshift \citep[e.g.,][]{Capak2015, Bouwens16, Bowler2018, Bethermin2020}. However, the cost to obtain statistical samples of galaxies in the EoR results in the fact that only a modest number of galaxies have been characterized in detail so far \citep[e.g.,][]{Watson2015, Smit2018, Laporte2019, Faisst2020, Harikane2021, Schouws2022}. Furthermore, the study of dust at $\mathrm{ 2 < z < 6}$ was for a long time limited to bright dusty galaxies such as submillimetre galaxies \citep[SMGs; e.g.,][]{Gruppioni2013, Wang2019a, Barrufet2020}. 
However, ALMA is bridging the gap between these extreme dusty massive galaxies and more moderate star-forming galaxies (see \citealt{Hodge2020} for a review). 

The recent observational improvements have allowed the discovery of the emergence of high-z dusty galaxies at $\mathrm{z > 6}$. In particular, \citet{Fudamoto2021} has serendipitously detected two dusty galaxies at $\mathrm{z_{spec} \sim 7}$ near massive neighbors at the same redshifts. This shows that dusty galaxies in the EoR could be more common than previously thought, which leads to the question of whether the number of dusty galaxies at $\mathrm{z>6}$ is higher than expected \citep[see also][]{Barrufet2022_JWST, Nelson2022, Rodighiero2022}. 

The possible underestimation of the number of dusty galaxies would have a direct impact on the obscured Star Formation Rate Density (SFRD), which remains uncertain at $\mathrm{z >3}$ \citep{Casey2019}. Several studies have calculated the obscured SFRD at $\mathrm{z>5}$ based on serendipitous sources resulting in largely differing conclusions  \citep[e.g][]{Gruppioni2020, Fudamoto2021, Talia2021, Casey2021, Viero2022}. 
While some studies find that 2mm selected, dusty galaxies contribute $\mathrm{\sim 30 \%}$ to the integrated star-formation rate density between $\mathrm{3 < z < 6}$ \citep{Casey2021}, others report a significantly larger obscured SFRD that remains constant over redshift \citep[e.g.,][]{Gruppioni2020, Talia2021}. An approach to clarify the contribution of dust-obscured star formation to the cosmic star formation history is to measure the infrared luminosity function (IRLF) all the way into the EoR. 
The shape and scale of the IRLF are crucial to understanding the abundance of dusty galaxies and how rapidly dust is formed in the early universe. This directly affects the fraction of star formation that is obscured in forming galaxies, and thereby the formation (or rise) of metals. 

Due to the wealth of rest-frame UV observations, the UV luminosity function (UVLF) is well constrained up to  $\mathrm{z \sim 9}$ \citep[e.g.,][]{Bouwens07, Bouwens15aLF, Oesch2018LF, Bowler2020, Bouwens2021a}, and we even have some information at $\mathrm{z \sim 9-10}$ \citep{Oesch2018, Harikane2022} and beyond now with JWST \citep[e.g.][]{Naidu2022_z11_z13, Donnan2022, Atek2022, Adams2022, Finkelstein2022}. 
In contrast, the IRLF is still quite uncertain at high redshifts. 
Current measurements of the IRLF rely on small numbers of dusty sources at $\mathrm{z > 3.5}$  \citep[e.g.,][]{Wang2019, Gruppioni2020}. 
This leads to large uncertainties in the IRLF parameters, including the faint-end slopes, and disagreements between different survey results \citep[e.g.,][]{Gruppioni2013, Koprowski2017, Lim2020, Popping2020, Gruppioni2020}. 

The recent study of \citet{Zavala2021} compiled the results of several surveys and combined those with semi-empirical modelling to constrain the evolution of the IRLF out to $\mathrm{z>5}$, albeit with significant uncertainties. However, an IRLF at $\mathrm{z \sim 7}$ has not been measured directly using dust continuum observations yet. In this context, we use the data from the Reionization Era Bright Emission Line Survey (REBELS), an ALMA large program aimed at obtaining a statistical sample of normal star-forming galaxies at $\mathrm{z > 6.4}$ (see \citealt{Bouwens2022} for details). REBELS has increased the number of spectroscopically observed massive galaxies in the EoR by a factor $\mathrm{\times \sim 4-5}$ compared to the previous literature \citep{Bouwens2021}. The same strategy of the REBELS selection was tested in a pilot program presented in \citet{Schouws2022}. This study showed the potential of ALMA as a high redshift `machine' and the six pilot galaxies are also included in the main REBELS sample \citep{Smit2018, Schouws2021, Schouws2022}. While observations from the REBELS program were just recently completed and analysis of the full data set now underway, its data have already been used for a number of scientific analyses, including the discovery of serendipitous dust-obscured sources at $z\sim7$ \citep{Fudamoto2021}, modelling the dust and ISM properties of $z>6$ galaxies  \citep[e.g.,][]{Sommovigo2022,Dayal2022,Ferrara2022}, measuring their detailed specific SFRs \citep{Topping22}, calculating their SFRD \citet{Algera2022}, estimating Ly$\alpha$ transmission around luminous sources in overdense $z\sim7$ environments \citep{Endsley2022}, and constraining the neutral gas fraction out to the EoR \citep{Heintz2022}.  

In this paper, we use this survey to calculate -- for the first time -- an IRLF at $\mathrm{z \sim 7}$. In Section \ref{observations}, we describe the ALMA observations and the infrared luminosity calculations used in this work. The methodology for calculating the IRLF and their values is described in Section \ref{Methodology}. We present the results on the obscured SFRD of REBELS galaxies in Section \ref{SFRD_section}. We discuss our results in Section \ref{Discussion} and present a summary and our conclusions in Section \ref{Summary}. 
 
\section{REBELS observations} 
\label{observations}

\subsection{ALMA observations and catalogue }
In this work, we use data from REBELS \citep{Bouwens2021} which is a Cycle 7 ALMA large program of $\mathrm{ \sim 40}$ UV bright galaxies at $\mathrm{z>6.4}$. The selection was based on UV brightness ($\mathrm{-23 < M_{UV} < -21.3}$) and photometric redshifts for galaxies identified over a combined area of $\mathrm{\sim 7 deg^{2}}$ in several fields (see \citealt{Bouwens2021} for details). This survey of spectral scan observations identifies bright ISM cooling lines ([CII], [OIII]) while simultaneously probing the dust-continuum in bands $\mathrm{158 \ \mu m }$ and $\mathrm{88 \ \mu m }$, respectively, which is essential to derive the infrared luminosity ($\mathrm{L_{IR}}$). Given its selection, the REBELS sample only spans a limited range in redshift and UV luminosities. 
Even though it is UV selected, the sample is representative of massive star-forming galaxies at $\mathrm{z \sim 7}$, providing an extensive probe of ISM reservoirs in the EoR \citep{Bouwens2022,Ferrara2022}. 

In this work, we only focus on galaxies that were scanned for [CII], i.e., sources with $\mathrm{z_{phot} = 6.4 - 7.7}$. The total sample used in this study contains 42 galaxies with [CII] scanned, 16 of which with a dust continuum detection at more than $3\sigma$. Notably, 15 of these 16 sources also do have a significant [CII] emission line detection and thus a robust spectroscopic redshift measurement \citep{Inami2022}.    

\subsection{Infrared luminosity from REBELS survey}
In this section, we describe the infrared luminosity measurements from \citet{Inami2022} and the average properties of the REBELS galaxies. 

When deriving the infrared luminosities of our sample, we have to make an assumption about the dust temperature. Estimating this based on a few photometric detections in the far-infrared is  very challenging. \citet{Sommovigo2021}  solve this difficulty using $\mathrm{L_{[CII]}}$ as a proxy for the dust mass and the underlying continuum to constrain the dust temperature. This is particularly useful for the REBELS survey, given that [CII] estimates (or upper limits) are available for the full sample. Using these measurements, \citet{Sommovigo2022}  find an average dust temperature of $\mathrm{T_{d}=46K}$ 
for the REBELS sample. Hence, \citet{Inami2022} assumed a Spectral Energy Distribution (SED) with dust temperature and emissivity from \citet{Sommovigo2022} ($\mathrm{T_{d}=46K}$ and $\mathrm{\beta =2}$ respectively) to calculate the infrared luminosity based on the ALMA dust continuum flux. For the galaxies without dust continuum detection a 
3$\sigma$ upper limit was derived both for the continuum flux and the corresponding infrared luminosity. A cosmic microwave background correction was applied for all galaxies, with and without dust detection. The correction depends on the exact redshift, but lies in the range of $\mathrm{8-14 \%}$ (see \citealt{Inami2022} for details). 

\begin{figure}
 \centering
   \includegraphics[width=\columnwidth]{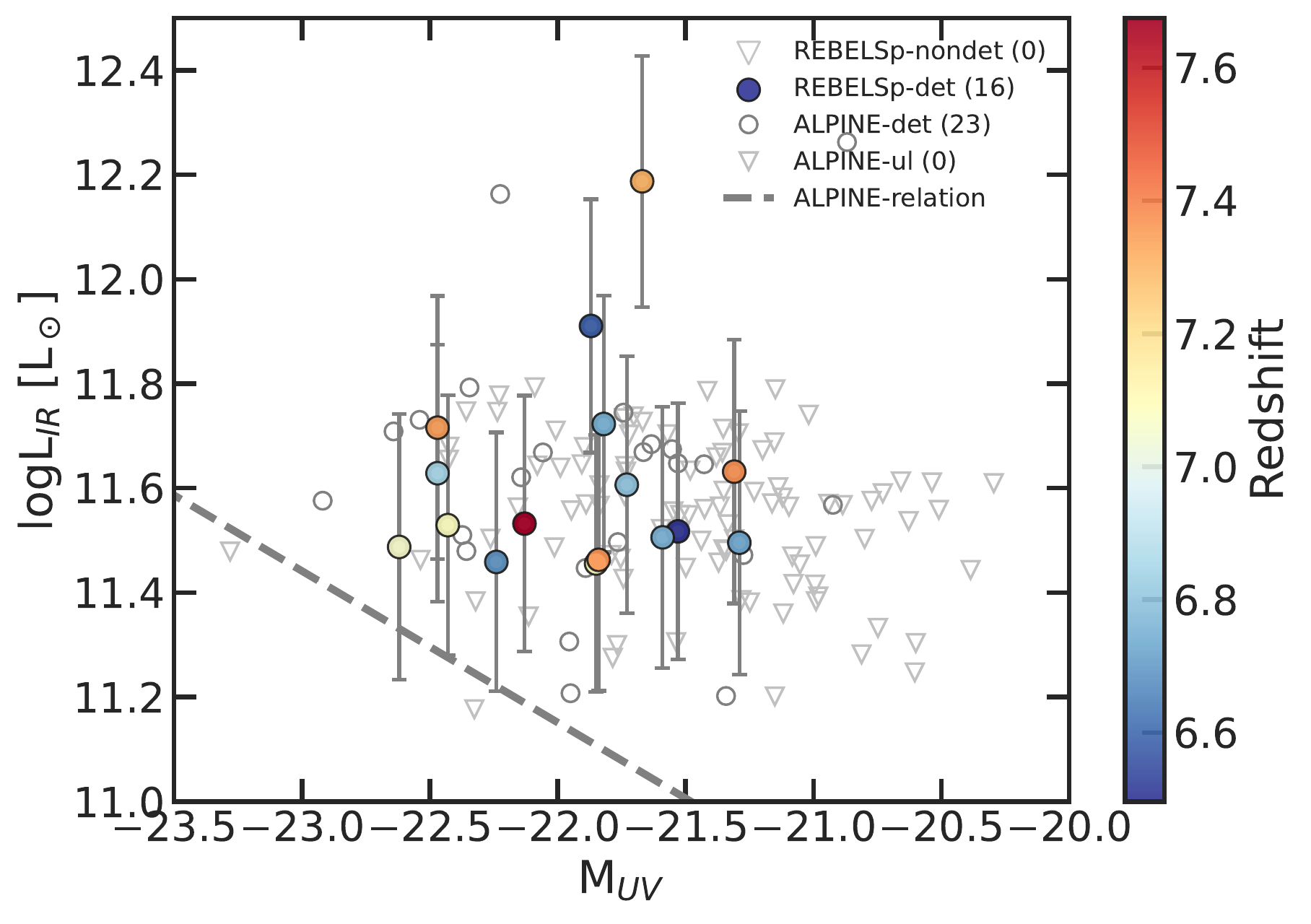}  
    \caption{ Infrared luminosity against UV absolute magnitude with the redshift colour-coded for the REBELS (filled symbols) and ALPINE (empty symbols) samples for both $\mathrm{3 \sigma}$ detections (dots) and upper limits (triangles). The REBELS sample does not show significant differences between detections and upper limits.  
 $\mathrm{L_{IR}}$ does not depend on $\mathrm{M_{UV}}$ or redshift. The small $\mathrm{L_{IR}}$ dynamic range and the flatness are comparable with the ALPINE sample at $\mathrm{4.5 < z < 6}$ although ALPINE extends to fainter UV galaxies (empty triangles and dots for upper limits and detections respectively). The ALPINE relation presented in \citet{Khusanova2021} is shown in the black dashed line. }
  \label{LIR_MUV}
\end{figure}

Using the derived IR luminosity measurements, we plot in Figure \ref{LIR_MUV} the relation between UV and IR-luminosities. Given the selection of UV luminous sources, the dynamic range both in UV and IR luminosities is limited. The REBELS sample only probes the most massive, UV-luminous galaxies at these redshifts. It is composed of luminous infrared galaxies (LIRGs; $\mathrm{10^{11} <  L_{IR}/ L_{\odot} < 10^{12} }$) except for REBELS-25, the brightest galaxy in our sample with $\mathrm{log(L_{IR}) \sim 12.2 L_{\odot}}$ (see Hygate et al. 2022 for details). 
The fact that we found only one ultra luminous infrared galaxy (ULIRG; $\mathrm{L_{IR} > 10^{12} L_{\odot} }$) in the REBELS sample could be due to the UV bright selection of REBELS galaxies with $\mathrm{-23 < M_{UV} < -21.3}$. We discuss this further in a later section. 

We compare the IR luminosities from REBELS with the sample from the ALMA Large Program to INvestigate [CII] at Early times \citep[ALPINE,][]{LeFevre2020} which targets UV-selected sources at lower redshifts at $\mathrm{4.5 < z < 6}$.  The ALPINE sample spans a wider $\mathrm{M_{UV}}$ range ($\mathrm{-23.3 < M_{UV} < -20}$) but is also mostly composed of LIRGs (see Figure \ref{LIR_MUV}) finding also in general dusty galaxies \citep[][Sommovigo et al. 2022b in prep]{Pozzi2021}. 
Our REBELS sample shows that UV-selected galaxies at $\mathrm{z \sim 7}$ have comparable infrared luminosities to UV-selected galaxies at lower redshift ($\mathrm{4.5 < z < 6}$) (see Section \ref{Discussion} for Discussion). 

\section{Infrared luminosity function at z $\sim$ 7} 
\label{Methodology}

In this section, we explain the procedure to calculate the luminosity function (LF). The main complication in computing a luminosity function using a targeted survey such as REBELS is that it is not straightforward to derive a selection volume for each source. This can be overcome by basing our volume estimates on the UV luminosity function as a proxy, as was successfully demonstrated in \citet{Yan2020} who used the ALPINE UV targeted sample to derive the [CII] luminosity function. Here, we closely follow their approach.

\subsection{Calculation of the luminosity function}
\begin{figure}
 \centering
   \includegraphics[width=\columnwidth]{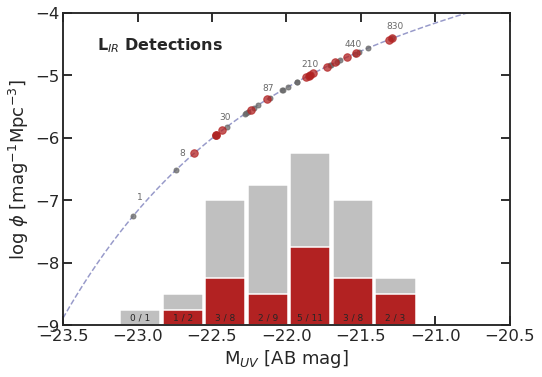}
  \caption{ Number of $\mathrm{L_{IR}}$
detections against the UV absolute magnitude.  
 The histogram shows the detected sources in red and the non-detections in grey with the fraction of detections/total indicated in the lower numbers.
 Also shown is the UVLF from \citep{Bouwens2021} as a dashed line. This is used to compute the representative volume for each of our targets. The small numbers above the LF indicate how many galaxies are expected per M$_{UV}$ bin in a volume-limited survey spanning the REBELS target selection area of 7 deg$^2$. Clearly, REBELS only targets a very small fraction of the full galaxy population at faint UV luminosities, which we account for in our analysis (see main text).
 } 
  \label{detections}
\end{figure}

Our derivation is based on the $\mathrm{z \sim 7}$ UVLF from \citet{Bouwens2021}. This is used to derive a representative volume for the UV-selected sources. In practice, we use the UVLF to compute the number of expected galaxies in bins of UV luminosity assuming a volume-limited survey over the full selection area of the REBELS sample of 7 deg$^2$ and $\mathrm{z=6.4-7.7}$ (see Fig. \ref{detections}). 
This is given by:
\begin{equation}
    \mathrm{N_{exp} = \phi_{UV}(M_{UV})\, \Delta M_{UV} \, V_{tot} }
\end{equation}
where $\mathrm{ \phi_{UV}(M)}$ is the UVLF from \citet{Bouwens2021} per magnitude bin $\mathrm{\Delta M_{UV}}$, and $\mathrm{V_{tot}}$ is the total survey volume over which REBELS sources were selected. REBELS only targets a very small sub-sample of all galaxies expected in such a large survey. We can compute a correction factor to account for this sampling incompleteness in each UV luminosity bin as $\mathrm{f_{UV}= N_{exp}/N_{obs}}$, where $\mathrm{N_{obs}}$ is the number of targeted REBELS galaxies in each $\mathrm{M_{UV}}$ bin. 

While the correction factor above is derived for a volume-limited survey, the requirement of a dust continuum detection can further introduce a reduction in the survey volume for each source. Namely, it can limit the maximum redshift up to which a given source would remain detected. This is accounted for by computing the so-called maximum comoving volume $\mathrm{V_{max,i}}$ for each galaxy i \citep[see][]{Schmidt68}. Specifically, $\mathrm{V_{max,i}} = \int_{z_{min}}^{z_{max,i}} \sfrac{d^2V}{dz\,d\Omega}\, \Omega\, dz$, where $\mathrm{z_{max,i}}$ is either the upper edge of the redshift bin of the LF, or, if smaller, the maximum redshift up to which source i would remain continuum detected at $>3\sigma$. $\Omega$ is the survey volume. In practice, $\mathrm{z_{max,i}}=7.7$ for most galaxies, except for the faintest few sources in the sample.

We now have all quantities to calculate the IR luminosity function $\mathrm{\phi_{IR}}$ in bins of $\mathrm{L{IR}}$. This is given by:
\begin{equation}
\mathrm{\phi_{IR}(\log L_{IR}) = \frac{1}{\Delta log L_{IR}} \sum_{i \in bin} \frac{f_{UV,i}}{V_{max,i} }  } \end{equation}
where i runs over all sources in a given IR luminosity bin $\mathrm{\log L_{IR}\pm\Delta \log L_{IR}/2}$ (see Eq. 3 in \citealt{Yan2020}). The uncertainties on the IRLF bins are computed as the Poisson errors in each $\mathrm{L_{IR}}$ bin. 

Note that this calculation is independent of the assumed survey area $\Omega$, since both $\mathrm{{V_{max}}}$ and $\mathrm{{f_{UV}}}$ are  directly proportional to it. 

We repeat the above calculation twice. In the first case, we only consider continuum detected galaxies (16 sources); in the second case, we include the full REBELS sample (42 sources), treating non-detections as upper-limits. The completeness factors $\mathrm{f_{UV}}$ are computed separately for both cases. The resulting IRLFs are in very good agreement, as discussed in the next section.

\subsection{The infrared luminosity function at $z \sim 7$}
\label{Results} 

\begin{figure*}
 \centering
  \includegraphics[width=1.0\textwidth]{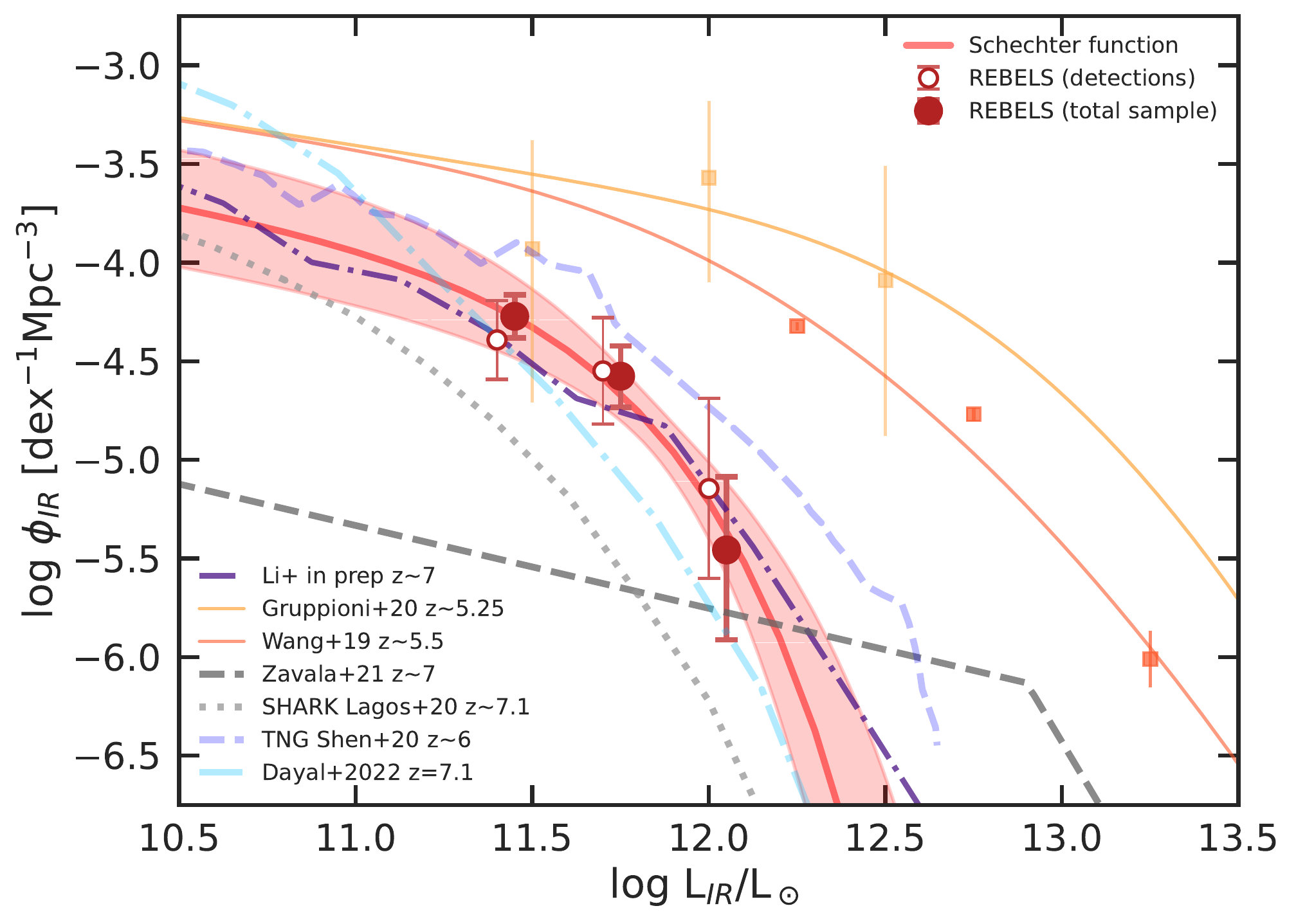} 
  \caption{
  Infrared luminosity function at $\mathrm{z \sim 7}$ for the REBELS sample (red dots and lines) compared with simulations (dashed lines) and observations (solid lines). The IRLF was calculated both only using the galaxies with dust continuum detections (16 galaxies, empty dots) as well as using the full sample including upper limits (42 galaxies, filled red dots). The red line shows the \citealt{Schechter1976} fit for the total sample. The shaded area shows the uncertainty of the luminosity function Schechter function fit with the total sample which is larger at the low luminosity end due to the lack of data.  
 The rest of the lines show both theoretical and observational IRLF studies in several fields. Our study is in agreement with Li et al. in prep (dark purple line) which predicts a similar number of dusty galaxies in a broad range of luminosities. The dark grey line is the IRLF at $\mathrm{z \sim 7}$ from \citet{Zavala2021} and predicts a larger number of galaxies than our study for the bright end with luminosities ($\mathrm{12.5 < log(L_{IR}/L_{\odot} < 13 }$)  whereas our luminosity function does not predict a significant number 
 of galaxies at $\mathrm{z \sim 7}$ with  $\mathrm{log(L_{IR}/L_{\odot} > 12.5 }$. 
 TNG simulations at $\mathrm{z \sim 6}$ from \citet{Shen2021} show a systematic shift with respect to our fitting, but consistent in shape (blue dashed line).  \citet{Dayal2022}
 and \citet{Lagos2020} simulations at $\mathrm{z \sim 7}$ (light blue and grey line respectively) present a 1 dex difference in the lower luminosity with our result in between them. The yellow line and dots indicate the IRLF at $\mathrm{z \sim 5.25}$ predicted by the serendipitous galaxies found in the ALPINE survey presented in \citet{Gruppioni2020}, whereas the orange symbols show \citet{Wang2019} results at similar redshift. }
  \label{IRLF}
\end{figure*}

\subsubsection{The Step-Wise IRLF}
In this section, we first present the step-wise LF  by using the methodology described in the previous subsection, before we derive parametric Schechter function fits.
Figure \ref{IRLF} shows the resulting LFs in 
three equidistant  luminosity bins $\mathrm{log(L_{IR}/L_{\odot}}$: [11.3-11.6], [11.6-11.9] and [11.9-12.2], both for our detections-only and our full sample. 
The derived stepwise LFs are in excellent agreement, showing that the detection-only sample is not biased significantly. In the rest of the paper, we use the total sample as a baseline.

For the detection-only sample, we further test the possible impact of uncertainties in the IR luminosity estimates. Specifically, we use a Monte Carlo technique in which we perturb the initial $L_{IR}$ measurements by their statistical (Gaussian) uncertainties 10,000 times and rederive the IRLF in each case. We then use the median and 16th and 84th percentiles, respectively, as the uncertainties.
We do not find significant differences in the resulting LF values, but the uncertainties are increased as can also be seen in Figure \ref{IRLF}.

\subsubsection{Schechter Function Fits}
We now derive a parametric estimate of the IRLF based on the classic Schechter function from \citealt{Schechter1976}, commonly used  both in the local and the high-z Universe \citep{Johnston2011}.  
The three parameters that define the Schechter function are $\mathrm{\phi^{*}}$, $\mathrm{L_{*}}$ and $\mathrm{\alpha}$; the normalization factor of the overall density of galaxies, the characteristic luminosity, and the faint-end luminosity slope, respectively. Due to the lack of data at low $\mathrm{L_{IR}}$, we have restricted $\mathrm{\alpha}$ taking into account the faint-end slope values found in the literature (see Section \ref{Discussion} for details). 
We fix the slope to $\mathrm{\alpha=-1.3}$ in our fitting, which is the value derived for the ALPINE high-z IRLF in \citet{Gruppioni2020}. 

We use a Bayesian  Monte Carlo Markov Chain (MCMC) approach to derive the posterior distribution of the Schechter function parameters. Hence, we compute the $\mathrm{\phi_{IR}}$, $\mathrm{L_{*}}$, while keeping the slope fixed at $\mathrm{\alpha=-1.3}$. 
We have set these initial parameters 
centered at the values obtained by minimizing the error function first ($\mathrm{log(\phi_{IR}) = -3.5}$, $\mathrm{log(L^{*}) = 11.7 }$), and then use non-informative Gaussian priors. 
We then perform 20,000 MCMC iterations and ensure that these are converged.  
We find that posterior distribution of the parameters is similar in both cases, either including the total sample (considering upper limits) or only detections. Therefore, we only present  the Schechter function with uncertainties for total sample in Figure \ref{IRLF}. 
The $\mathrm{1\sigma}$ uncertainty of the fit function was also calculated from the MCMC chains computing the 16th and 84th percentiles of the posterior distributions. The $\mathrm{\phi_{IR}}$ uncertainties in the fainter end are $\mathrm{\sim 0.5 \ dex}$, while at the brighter end they are $\mathrm{<0.2 \ dex}$. 
The IRLF is best constrained between $\mathrm{ 11.5 < log(L_{IR}/L_{\odot}) < 12}$, and shows that the density of sources drops quickly ($\mathrm{log(\phi_{IR})< -6.5 dex^{-1} Mpc^{-3} }$) at luminosities above $\mathrm{ log(L_{IR}/L_{\odot}) > 12.3  }$. 

The resulting Schechter function parameters are $\mathrm{log (\phi_{IR})= -4.38^{+0.38}_{-0.35} dex^{-1} Mpc^{-3} }$ and  $\mathrm{log(L_{*}/L_{\odot}) = 11.60^{+0.23}_{-0.13}}$ with a fixed $\mathrm{\alpha =-1.3}$ (see Table \ref{summarizeTable} for the summary of the main parameters). 
Our analysis shows a $\mathrm{z \sim 7}$ IRLF with a considerable number of LIRGs that drops in the ULIRG range suggesting a limit in luminosity at  $\mathrm{log(L_{IR}/L_{\odot}) \sim 12.3}$. 
This is in general agreement with some theoretical studies. The IRLF at $\mathrm{L_{IR} < 11.5 L_{\odot}}$ is uncertain and a larger study with fainter galaxies should be carried out to accurately measure the IRLF at the fainter luminosity end. 

We compared our results to both theoretical and observational IRLF studies at similar redshifts (see dashed and continuous lines respectively in Figure \ref{IRLF}). Generally, our results are in broad agreement with some simulated IRLFs at similar redshift. When comparing to lower redshift observations at $z\sim5-6$, however, we find that our IRLF is more than an order of magnitude lower. Finally, our IRLF shows an interesting evolution with redshift, compared with the literature, not only in number density (as was previously shown in \citet{Koprowski2020, Fujimoto2023}), but also in $\mathrm{L_{*}}$. This could be due to our UV-selected sample being biased to bright sources and further study with a similar selection at different redshift should be carried out to confirm the possibility of evolution with $\mathrm{L_{*}}$. 

We discuss the points above in more detail in Section \ref{Discussion}. We also discuss in subsection \ref{caveats} the importance that our data is UV-bright selected which cannot take into account extremely dust-obscured sources that are faint in the UV.


\section{Obscured star formation rate density}
\label{SFRD_section}

In this section, we calculate the obscured SFRD directly through the IRLF derived in the previous section. We calculate the SFRD in two different ways: 1) by simply summing up the step-wise infrared densities for the data in the REBELS sample and 2) by integrating the Schechter IRLF over the luminosity range  $\mathrm{ 10.5 < log(L_{IR}/L_{\odot}) <13}$. These limits were selected in the range over which we can define the Schechter function. Note that the integration limits are narrow but, due to the luminosity bins, there is no data to constrain a lower-limit integration. Further analysis is produced in section \ref{Discussion}. In both cases we use a conversation factor $\mathrm{\kappa = 10^{-10} M_{\odot}/yr/L_{\odot}}$.

For the step-wise estimates, we considered both the total sample and detections. 
We find $\mathrm{log(SFRD/(M_{\odot}/yr/Mpc^{3})) = -3.21 \pm 0.18 }$ taking only into account the dust continuum detections, which is slightly lower than for the total sample with $\mathrm{log(SFRD/M_{\odot}/yr/Mpc^{3}) = -2.93 \pm 0.20 }$. This SFRD estimate needs to be considered as a lower limit, since it only takes into account the three luminosity bins. 

\begin{figure*}
 \centering
  \includegraphics[width=\textwidth]{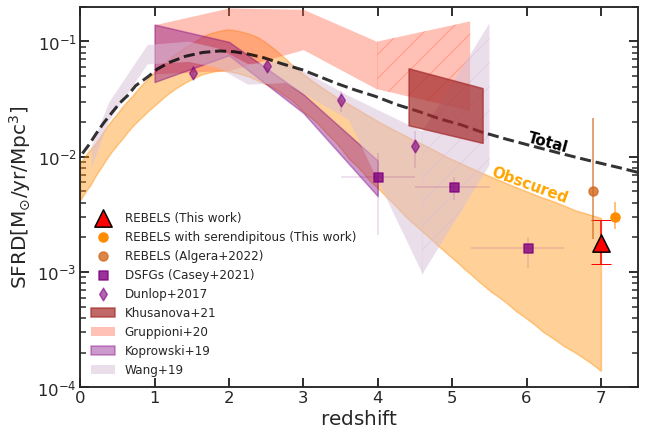}
   \caption{ Star formation rate density against redshift for the REBELS sample at $\mathrm{z \sim 7}$ and several works in the literature. The black line shows the total SFRD from \citet{Madau2014} whereas the orange shaded region shows the obscured SFRD \citep{Zavala2021}. Our results show a moderate SFRD calculated from  the fitted IRLF (red triangle) which increases if the two serendipitous normal dusty REBELS galaxies from \citet{Fudamoto2021} are taken into account (orange dot). Similarly, \citet{Algera2022} obtains a larger contribution to the obscured star formation but in agreement within $\mathrm{1 \sigma}$ error (dark orange dot). 
   DSFGs from the ALMA 2 mm photometric blind survey show a decrease in SFRD over redshift (purple squares; \citealp{Casey2021}). The 1.3 mm ALMA blind survey presented in \citet{Dunlop2017} shows a obscured SFRD at $\mathrm{1 < z < 4.5}$ that decreases at $\mathrm{z>2}$ (purple diamonds).
   \citet{Khusanova2021} shows the SFRD from the ALPINE survey at $\mathrm{z \sim 5}$ (brown area). Also from ALPINE, \citealp{Gruppioni2020} present a larger obscured SFRD which is decreasing at $\mathrm{z>3}$ (pink area) with the last redshift bin at $\mathrm{z>4}$ containing only one source (dashed pink area). Similarly, \citet{Wang2019} shows a decreasing SFRD (light purple area) with large uncertainty in the last bin at $\mathrm{z \sim 4}$ (dashed light purple area). \citet{Koprowski2020} presented a constrained SFRD up to $\mathrm{z \sim 4}$ (purple area). REBELS results shows the presence of dust at $\mathrm{z \sim 7 }$ even in UV-selected galaxies. }
  \label{SFRD}
\end{figure*}

To extrapolate to fainter luminosities, we have calculated the SFRD for the Schechter LFs. In particular, we use the MCMC chains to derive the median posterior SFRD and the associated uncertainties.
We find $\mathrm{log(SFRD/M_{\odot}/yr/Mpc^{3}) = -2.66^{+0.17}_{-0.14} }$ where the uncertainties correspond to the 16-84th percentile (see Figure \ref{SFRD} ). As expected, this SFRD is larger than the SFRD calculated from the observations, since it is integrated over the full luminosity range ( $\mathrm{10.5 <log(L_{IR}/L_{\odot}) <13}$). Notice that REBELS is a UV-selected sample and the obscured SFRD needs to be taken into account as a robust lower limit (see caveats in Section \ref{caveats}). Finally, the SFRD was computed adding the serendipitous sources from the REBELS sample presented in \citet{Fudamoto2021}. The sum of the two points, UV-selected galaxies and serendipitous 'dark' systems, is
$\mathrm{log(SFRD/(M_{\odot}/yr/Mpc^{3})) = -2.53 ^{+0.17}_{-0.14} }$. 

\begin{table}
 \centering
 \begin{tabular}{|c|c|c|c|}
 $\mathrm{\alpha}$  & $\mathrm{log(L^{*})}$ & $\mathrm{log(\phi_{IR})}$ &  $\mathrm{log(SFRD)}$   \\
   & $\mathrm{[L_{\odot}]}$ & $\mathrm{[dex^{-1} Mpc^{-3}]}$ &  $\mathrm{[M_{\odot}/yr/Mpc^{3}]}$   \\
 \hline 
    \multicolumn{4}{c}{Schechter Function Fit} \\
 -1.3 (fix) 	& $\mathrm{11.60}^{+0.23}_{-0.13}$    & $\mathrm{-4.38^{+0.38}_{-0.35}}$ &   $\mathrm{ -2.66^{+0.17}_{-0.14}}$  \\
 \hline
 Total sample   & $\mathrm{11.15}$ 	& $\mathrm{-4.3^{+0.1}_{-0.1}}$ & $\mathrm{-2.93 \pm 0.20 }$    \\
  & $\mathrm{11.75}$ & $\mathrm{-4.6^{+0.2}_{-0.2}}$ &     \\
  & $\mathrm{12.05}$ & $\mathrm{-5.5^{+0.4}_{-0.5}}$ &     \\ \hline
 Detections   & $\mathrm{11.15}$ 	& $\mathrm{-4.4^{+0.2}_{-0.2}}$  &   $\mathrm{-3.21 \pm 0.18 }$  \\
 & $\mathrm{11.75}$ 	& $\mathrm{-4.6^{+0.3}_{-0.3}}$  &     \\
   & $\mathrm{12.05}$ 	&  $\mathrm{-5.1^{+0.2}_{-0.5}}$ &     \\ \hline
    \end{tabular}       
        \caption{Summary of the main parameters of this study. The first column shows the faint luminosity slope ($\mathrm{\alpha}$), and the second column shows the luminosity function at the determined luminosity bin (third column). Finally, the fourth column shows the obscured star formation rate density taking into account the three luminosity bins. The first row shows the best fit Schechter function parameters for a fixed slope of $\mathrm{\alpha = -1.3}$, while the subsequent rows show the total sample and only with detections. }  
         \label{summarizeTable} 
        \end{table} 
        
We compare our results with previous studies in the literature for both similar samples  to REBELS and other dusty galaxies at high redshift. Our derived obscured SFRD of the REBELS sample is $\mathrm{\sim 13\pm 1 \% }$ of the total CSFRD at $\mathrm{z \sim 7}$ from \citet{Madau2014} 
and $\mathrm{9 \% }$ of the unobscured SFRD estimate from \citet{Bouwens2022}. 
This is in agreement with the range of obscured SFRD predictions of \citet{Zavala2021}, who use a compilation of several surveys to derive a model of the IRLF evolution. Our resulting obscured SFRD lies in the upper part of their inferred SFRD range being the first result at $\mathrm{z \sim 7}$ calculated through [CII] spectroscopic scans. In an accompanying paper, \citet{Algera2022} also derived the SFRD for the REBELS sample using the stellar mass as a proxy to calculate the SFRD through a stacking analysis. While our best estimates are a factor $\sim2.5\times$ lower, 
the measurements are consistent within the $\mathrm{1 \sigma}$ uncertainties.

In Figure \ref{SFRD} we also present the obscured SFRD for several studies showing the lack of consensus at $\mathrm{z> 3}$ on the obscured SFRD.
Our SFRD result is comparable to DSFGs selected at 2mm from \citet{Casey2021}, who reports a decrease in the  obscured SFRD over $\mathrm{4 <  z < 6}$.  In contrast to these findings, the SFRD from serendipitous sources found in the ALPINE survey present a non-evolving SFRD  across the whole redshift range of the sample ($\mathrm{1 < z < 5.5}$). Their calculated SFRD is over two orders of magnitude more than our results at $\mathrm{z\sim7}$. Similarly, longer wavelength studies support a flatter evolution of the SFRD at $\mathrm{ 3 < z <6}$, albeit with more moderate SFRD \citep{Talia2021}.
In contrast, our results show lower SFRD at $\mathrm{z \sim 7}$, which, when compared to literature at lower redshifts, supports a non-flat SFRD across redshift (see section \ref{Discussion} for discussion). 

\section{Discussion}
\label{Discussion} 

In this section, we compare our IRLF results with observational and theoretical studies. However, due to the underlying assumptions, IRLFs from simulations are not directly comparable. As a result, our findings broadly concur with theoretical research. On the observational side, the literature shows a large range of IRLF suggesting SFRD discrepancies of $\mathrm{ \sim 2}$ orders of magnitude. We also explore the causes for the different results in the literature and compare to our IRLF and SFRD.

\subsection{Comparison to Literature}

Some theoretical IRLFs at $\mathrm{z \sim 6-7}$ agree quite well with our findings. For example, Li et al. in prep. show a similar IRLF over the luminosity range $\mathrm{ 10.5 < log(L_{IR}/L_{\odot}) < 12.5}$, as do the TNG+300 simulations shown in \citet{Shen2021}. But throughout the whole infrared luminosity range, the latter exhibits larger number densities by $\mathrm{\sim 0.5 dex}$. A plausible explanation for this shift is the difference in redshift ($\mathrm{\Delta z \sim 1}$) between our results and those of \citet{Shen2021}, as the IRLF is expected to decrease in number density at increasing redshift (see e.g. \citep{Koprowski2017, Fujimoto2023}). 

Our results contrast with those from \citet{Lagos2020} 
which themselves differ by $\mathrm{ \sim 0.5 dex}$ despite the fact that both utilise semi-analytical models based on merger trees. Over the full range of our directly observed luminosities ($\mathrm{log(L_{IR}/L_{\odot})> 11.5}$), our results are higher than both of these estimates.

Although the simulations described above are based on different assumptions, the theoretical work does not contain a UV selected sample bias. This suggests that, according to simulations, our IRLF estimate is not missing a significant number of extremely luminous, UV-undetected galaxies at $\mathrm{z \sim 7}$ (for potential caveats, see Section \ref{caveats}).  

We continue by contrasting with semi-empirical models from \citet{Zavala2021} at $\mathrm{z \sim 7}$. 
Their IRLF changes very little at $\mathrm{12< log(L_{IR}/L_{\odot}) < 12.5} $, whereas our IRLF sharply declines.  Our study shows an IRLF an order of magnitude higher for LIRGs and a negligible number of galaxies with $\mathrm{log(L_{IR}/L_{\odot}) > 12.3}$. Thus, we find a different distribution also for the bright luminosity end. These differences in IRLF could be explained by the different methodology, due to the lack of observational data at $\mathrm{z \sim 7}$, that leads to an extrapolation of their IRLF at higher redshifts. To do that, it is necessary to assume two different slopes for the LIRGs and the ULIRGs that might lead to different outcomes  between our study and \citet{Zavala2021}. 

Finally, we compare our results with IRLFs derived from observations. In particular, we contrast with the ALPINE IRLF, since it is an analogous survey to REBELS, but at lower redshift (see section \ref{observations} for details). Using the ALPINE data, \citet{Gruppioni2020} provide the IRLF at $\mathrm{z \sim 5}$ for serendipitous galaxies. Their IRLF agrees with ours for the lower luminosity bin, but the overall normalisation is significantly higher. The reason of the difference is the IRLF rely on several factors. Firstly, the redshift difference ($\mathrm{\Delta z \sim 2}$) is an obvious reason for the density to be lower. Furthermore, the REBELS sample was UV-selected, implying a selection effect that is nonexistent in a blind survey (see section \ref{caveats} for caveats). Another cause for the disparity with \citet{Gruppioni2020} might the difference it redshift calculation. Their redshifts were calculated with multi-band photometry and with only three galaxies at $\mathrm{z \sim 5}$. Finally, the differing dust temperature assumptions and the SED fitting may lead to different infrared luminosities, but further analysis is required to ensure that the differences are significant.

In order to continue the observational comparison, we contrast the IRLF calculated with the maximum redshift observed to yet in \citet{Wang2019}.
This analysis presents an IRLF with bright infrared galaxies selected with Herschel Space Observatory \citet{Pilbratt2010} at $\mathrm{z=5.5}$. 
At same redshift, their results have a 2 dex greater luminosity function than ours at the bright end, but a smaller overall luminosity function than the one stated in \citet{Gruppioni2020}. Again, the expected difference is caused by the disparity in redshift, as does the bias to select massive galaxies with Herschel. 

\subsection{What IRLF is needed to reproduce extreme SFRD?}

This section discusses how changes in the IRLF impact the SFRD. Since there is lack of consensus about obscured SFRDs at $\mathrm{z>5}$, we evaluate the key variables that influence  the SFRD computation: the IRLF faint end slope, the $\mathrm{L_{IR}}$ integration limits, and the conversion factor between $\mathrm{L_{IR}}$ and SFRD. To do that, we compute the SFRD derived for extreme $\mathrm{\alpha}$ and  integration limits to determine whether the most extreme SFRD described in the literature could be reproduced.  We also discuss the likely causes of these variances.

\begin{figure}
 \centering
   \includegraphics[width=1.0\columnwidth]{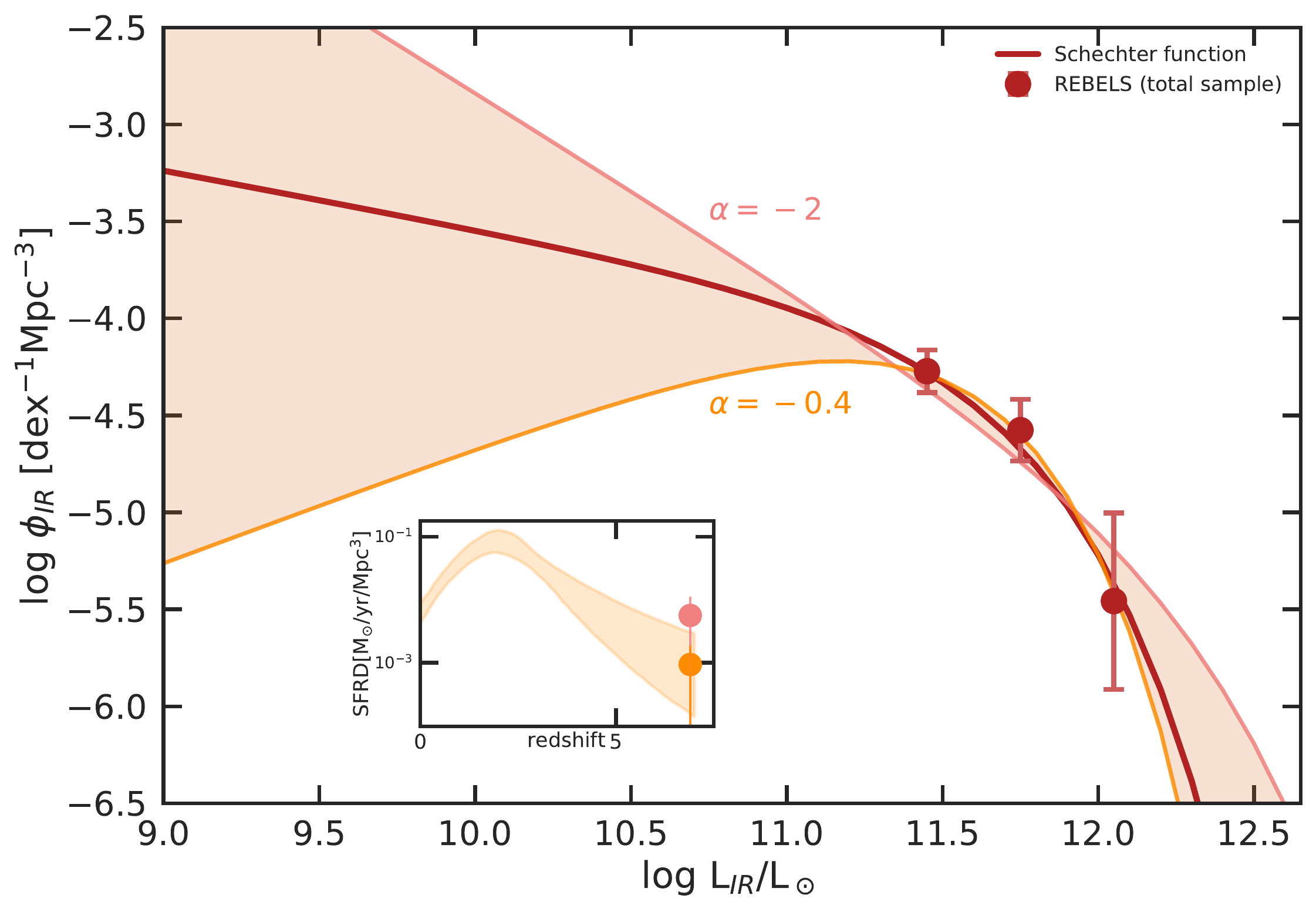}
 \caption{ The SFRD depends on the IRLF shape and the luminosity range used in the integration. The faint end slope $\alpha$ assumed in the low luminosity end is key for the resulting SFRD. This plot shows the best fit IRLF for two extreme slopes:  $\mathrm{\alpha=-2}$ (red line) and $\mathrm{\alpha=-0.4}$ (orange line). The difference between slopes increases in IRLF being $\mathrm{\sim 4}$ orders of magnitude higher at $\mathrm{L_{IR} = 10^{9}L_{\odot} }$. The inner plot shows the SFRD for these two extreme cases which shows an order of magnitude difference depending on the slope assumed with the same integration luminosity ($\mathrm{10^{8} < L_{IR}/L_{\odot} < 10^{13}}$). 
The dark red dots show the total REBELS sample  
for the three luminosity bins. The dark red line shows the Schechter fit with $\mathrm{\alpha=-1.3}$ (dark red line) as presented previously in Section \ref{IRLF}. }
  \label{IRLF_severalalphas}
\end{figure}

First, we investigate changes in the IRLF slope.
Lower redshift studies frequently find a slope of $\mathrm{\alpha=-1.3}$, including more galaxies with lower infrared luminosities \citep{Hammer2012}, but some high redshift studies report shallower faint-end slopes of $\mathrm{\alpha=-0.4}$ \citep{Koprowski2017, Zavala2021}. 
In Figure \ref{IRLF_severalalphas}, we compute the IRLF for these two extreme cases by using $\mathrm{\alpha = -2}$ and $\mathrm{\alpha = -0.4}$, respectively. 
Additionally, we used a wider luminosity range for the  integration than in previous sections of this work, allowing for $\mathrm{8 < log(L_{IR}/L_{\odot}) < 13}$ as in \citep{Gruppioni2020}. Nevertheless, we cannot recreate values close to their SFRD, even in the most extreme scenario ($\mathrm{\alpha=-2}$), yielding a $\mathrm{SFRD \sim 6 \cdot 10^{-3} M_{\odot}/Mpc^{3}/yr}$. 

This SFRD is, however, consistent with the findings of \citet{Talia2021} ($\mathrm{SFRD \sim 5 \cdot 10^{-3}  M_{\odot}/yr/Mpc^{3} }$ at $\mathrm{z \sim 5}$).
It should be noted that the analysis of \citet{Talia2021}  was conducted using radio galaxies with median $\mathrm{ L_{IR} = 2.3 \pm 0.5 \times 10^{12} L_{\odot} }$, and is thus based on a different set of assumptions than our IR-based estimates. 

Despite the fact that it is common to compute the obscured SFRD using the IRLF, some studies directly calculate it by using the individual SFRs.  For instance, the MORA survey  performed blind 2mm ALMA observations \citep{Casey2021}, and identified a number of $\mathrm{z\sim4-6}$ DSFGs.  They find $\mathrm{SFRD \sim 10^{-3} \  M_{\odot}/yr/Mpc^{3}}$ at $\mathrm{z \sim 6}$, which is far lower than the previously mentioned studies such as \citet{Talia2021} or \citet{Gruppioni2020}.
The key distinction is that their photometric redshift estimates are based on submillimetre data, which can be degenerate with dust temperature. Generally, however, the findings of \citet{Casey2021} are in good agreement with ours, and their obscured SFRD is compatible with a $\mathrm{ z \sim 6}$ extension of our SFRD at $\mathrm {z \sim 7}$. This agreement also extends to the 1.3 mm ALMA serendipitous sources at $\mathrm{z < 4.5}$ from \citet{Dunlop2017}. Both \citet{Dunlop2017} and \citet{Casey2021} present a decrease of obscured SFRD at $\mathrm{z>3}$ which likely continues beyond $\mathrm {z >6}$, as suggested by our data. 

Even if several obscured SFRD present large values at $\mathrm{z \sim 5}$ (i.e. \citet{Wang2019, Gruppioni2020, Khusanova2021}), we also notice that the highest redshift bin in both \citet{Wang2019} and \citet{Gruppioni2020} have larger uncertainty than the rest to the low number of sources (as shown the hatched areas in Figure \ref{SFRD}). Given these larger uncertainties, a declining SFRD cannot be excluded from these analyses. Hence, although not in agreement, our results are not in contradiction with the studies that show large SFRDs and the highest redshift surveys. Studies including larger samples at $\mathrm{4 < z < 7}$ would be needed to corroborate this hypothesis.

\subsection{Possible Caveats}
\label{caveats}

In this section, we assess the importance of our data being based on a UV-bright target selection. This directly implies that our study cannot account for extremely dust-obscured sources, such as SMGs, that are faint in the UV. However, given that there are several verified SMGs at $\mathrm{ z > 4}$, we know that such galaxies are 100$\times$ less common than UV-based Lyman Break Galaxies, given the SMG sky surface density of 0.01 arcmin$^{-2}$ \citep[e.g.,][]{Riechers2013, Riechers2017, Marrone2018}. Furthermore, extremely dusty high redshift galaxies have only been discovered up to a maximum $\mathrm{z = 6.34}$ \citep{Riechers2013}. All of these findings are based on large surveys conducted with the South Pole Telescope (SPT), SCUBA-2, or Herschel Space Observatory. 

The serendipitous detection of two dust-obscured galaxies in the REBELS dataset with similar masses and SFRs as the main sample clearly shows that the primary target sample of REBELS is not complete \citep{Fudamoto2021}. While, the  contribution of this class of galaxies to the SFRD is still very uncertain, \citet{Fudamoto2021} estimate a value of $\mathrm{1.2\times10^{-3}\,M_{\odot}/yr/Mpc^{3}}$, i.e. comparable to our estimate from the IRLF. This would suggest that UV-undetected galaxies could contribute a similar, but additional amount of obscured SFR as UV-bright galaxies. 

Similar conclusions have been reached from recent JWST observations. The first deep NIRCam observations revealed the existence of UV-undetected, dusty galaxies at $\mathrm{z>6}$. 
\citet{Barrufet2022_JWST}, in particular, present the SFRD for high-z dusty galaxies, finding a $\mathrm{log(SFRD/M_{\odot}/yr/Mpc^{3}) \sim -3}$ at $\mathrm {z \sim 7}$ for highly attenuated galaxies. 
We thus conclude that the galaxies we are missing in UV selections might contribute the same order of magnitude as the REBELS sample itself. 

To compute a more complete IRLF it would be necessary to perform a deep, but blind survey to probe galaxies at $z\sim7$ at several wavelengths. For the present, a good first step is to obtain results based on the UV-selected REBELS galaxies.  These results
represent a firm lower limit on the total obscured SFRD at $\mathrm{z \sim 7}$. 

\section{Summary and Conclusions} 
\label{Summary}

In this work, we have exploited the data from the REBELS survey, which consists of ALMA spectroscopic data of UV-bright galaxies in the EoR. 
Our sample consists of 42 galaxies at $\mathrm{6.4 < z < 7.7}$. 16 have revealed significant dust continuum emission at rest-frame $\sim158\micron$, and all but one of these are spectroscopically confirmed through their [CII] emission lines. This sample was used to:

\begin{itemize}
   \item We have calculated the Infrared Luminosity Function (IRLF) at $\mathrm{z \sim 7}$ for the first time using a spectroscopically confirmed sample. We find a $\mathrm{log (\phi_{IR}) \sim -4.2 \pm 0.2 \ dex^{-1} Mpc^{-3} }$ in our faintest luminosity bin of $\mathrm{log(L_{IR}/L_{\odot}) \sim 11.5}$. At higher luminosities, the IRLF decreases considerably. 
   
    \item We have fit a \citet{Schechter1976} function with a fix slope of $\mathrm{\alpha = -1.3}$ for the low luminosity end finding the best fitting values $\mathrm{log (\phi_{IR}) \sim -4.38 \ dex^{-1} Mpc^{-3} }$ and $\mathrm{log(L_{IR}/L_{\odot}) =11.6}$. 
    Our results indicate that extremely luminous galaxies with $\mathrm{log(L_{IR}/L_{\odot}) > 12.3}$ are extremely rare at $z\sim7$, with number densities  $\mathrm{log(\phi_{IR}) < -6.5 dex^{-1} Mpc^{-3}}$.

    \item We have derived the obscured Star Formation Rate Density through the IRLF. From the observations we calculate a lower limit of $\mathrm{log(SFRD/M_{\odot}/yr/Mpc^{3}) = -2.93 \pm 0.20 }$ at $\mathrm{z \sim 7}$ which represents $\mathrm{\sim 13 \%}$ of the total SFRD. When integrating over the luminosity range $\mathrm{10.5 < log(L_{IR}/L_{\odot}) < 13}$ we infer a larger value of $\mathrm{log(SFRD/M_{\odot}/yr/Mpc^{3}) = -2.66^{+0.17}_{-0.14}}$. 
    
    \item Our IRLF is broadly consistent  with  some simulations at $\mathrm{z \sim 7}$. The inferred SFRD is a robust lower limit that shows a significant contribution of obscured star formation at $\mathrm{z \sim 7}$. 
    
\end{itemize}

We conclude that our results imply a significant amount of obscured SFR at $\mathrm{z \sim 7}$ of at least $\mathrm{log(SFRD/M_{\odot}/yr/Mpc^{3}) \sim -3}$. Comparing with ALMA blind surveys, our results suggest a steep evolution of the obscured SFRD over redshift that continues to $\mathrm{z \sim 7}$, at least.

\section*{Acknowledgements}

We acknowledge the constructive feedback of the referee (MB) for his constructive feedback that helped in the improvement of this paper. We acknowledge support from: the Swiss National Science Foundation through the SNSF Professorship grant 190079 (LB and PAO). 
The Cosmic Dawn Center (DAWN) is funded by the Danish National Research Foundation under grant No.\ 140.
PD acknowledges support from the European Research Council's starting grant ERC StG-717001 (``DELPHI"), from the NWO grant 016.VIDI.189.162 (``ODIN") and the European Commission's and University of Groningen's CO-FUND Rosalind Franklin program.
AF and AP acknowledge support from the ERC Advanced Grant INTERSTELLAR H2020/740120. Generous support from the Carl Friedrich von Siemens-Forschungspreis der Alexander von Humboldt-Stiftung Research Award is kindly acknowledged. 
YF acknowledges support from NAOJ ALMA Scientific Research Grant number 2020-16B.
VG gratefully acknowledges support by the ANID BASAL projects ACE210002 and FB210003.



\bibliographystyle{mnras}
\bibliography{MasterBiblio} 









\bsp	
\label{lastpage}
\end{document}